\documentclass[reprint,amsmath,amssymb,aps]{revtex4-2}

\usepackage{graphicx}
\usepackage{tikz}
\usepackage{float}

\usepackage{amsmath}
\usepackage{amsfonts}
\usepackage{amssymb}
\usepackage{bm}

\usepackage{dcolumn}

\usepackage{color}
\usepackage{xcolor}
\usepackage{ulem}

\usepackage{enumerate}

\usepackage[colorlinks=true,linkcolor=blue,urlcolor=blue,filecolor=black,citecolor=blue,pdfstartview=FitV,pdftitle={},pdfsubject={},pdfkeywords={},pdfpagemode=None,bookmarksopen=true]{hyperref}

\begin{document}
\title{
	Capturing quantum phase transition in the ultraviolet region by holography
}

\author{Fang-Jing Cheng $^{1,2}$}
\email{fjcheng@mail.bnu.edu.cn}
\author{Zhe Yang $^{1,3}$}
\email{yzar55@stu2021.jnu.edu.cn}
\author{Yi Ling $^{4,5}$}
\email{lingy@ihep.ac.cn, corresponding author}
\author{Jian-Pin Wu $^{3}$}
\email{jianpinwu@yzu.edu.cn}
\author{Zhou-Jian Cao $^{2,6}$}
\email{zjcao@amt.ac.cn}
\author{Peng Liu $^{1}$}
\email{phylp@email.jnu.edu.cn, corresponding author}

\affiliation{
	$^1$ Department of Physics and Siyuan Laboratory, Jinan University,           Guangzhou 510632, China\\
	$^2$ School of Physics and Astronomy, Beijing Normal University, Beijing, 100875, China\\
	$^3$ Center for Gravitation and Cosmology, College of Physical Science and Technology, Yangzhou University, Yangzhou 225009, China\\
	$^4$ Institute of High Energy Physics, Chinese Academy of Sciences,           Beijing 100049, China\\
	$^5$ School of Physics, University of Chinese Academy of Sciences,            Beijing 100049, China\\
	$^6$ Institute for Frontiers in Astronomy and Astrophysics, Beijing Normal University, Beijing 102206, China
}

\begin{abstract}

	We reveal for the first time that ultraviolet (UV) observables can diagnose quantum phase transitions (QPTs). In a class of holographic models exhibiting metal-insulator transitions, we study two types of UV observables: high-frequency conductivity and short-range entanglement. Remarkably, we find that the derivatives of these UV observables exhibit extrema near the quantum critical point. Analytical results show these critical behaviors arise from the deformation of the asymptotic bulk geometry. Moreover, these UV diagnostics show enhanced robustness to thermal fluctuations compared to typical infrared (IR) diagnostics, providing a clean method to identify quantum criticality at finite temperature. This work opens a new window for exploring quantum critical phenomena, suggesting that high-energy spectroscopic or entanglement-based probes could offer a robust alternative to traditional low-energy transport measurements.

\end{abstract}
\maketitle

{\it Introduction.---}
Quantum phase transitions (QPTs) are prominent phenomena in strongly correlated systems, conventionally characterized by the critical point separating distinct ground states~\cite{Sachdev:2011fcc,Donos:2012js}. Diagnosing the occurrence of the QPT by locating the critical point at the zero temperature is a fundamental problem in quantum critical phenomenon. Usually, the infrared (IR) fixed point of the renormalization group (RG) flow  dominates their universal characteristics; consequently, traditional studies have primarily focused on IR physics~\cite{Fisher:1990ux,Vidal:2002rm,Osterloh:2002aeb}. However, diagnosing QPTs solely through IR observables faces two key challenges. (i) Many IR observables are highly susceptible to thermal smearing, limiting their effectiveness in diagnosing QPTs~\cite{Basov:2011zz}. (ii) They are insensitive to high-energy features, failing to capture microscopic signatures that may also encode critical information~\cite{Sachdev:2011fcc}.

Recent advances in both theory and experiment provide a novel approach to tackling these challenges. On the theoretical side, certain strongly correlated systems show that high-energy (UV) degrees of freedom influence low-energy (IR) scaling behavior~\cite{Seiberg:1999vs,deBoer:1999tgo}, which implies that the information of quantum criticality, typically associated with IR dynamics, may also be encoded in the UV regime, where universal dynamics for high-momentum modes are predicted~\cite{Khlebnikov:2019yld}. Moreover, the inherently high energy scale of UV physics far exceeds typical thermal fluctuations. This property suggests that UV observables are more insensitive to thermal effects, making them powerful candidates for diagnosing QPTs at finite temperature. On the experimental side, UV-sensitive techniques such as high-harmonic generation have recently demonstrated the ability to diagnose a QPT~\cite{Heide:2022}. Nevertheless, it remains unclear whether the quantum criticality leaves any detectable signature in the UV region of the strongly correlated systems. One major obstacle results from the exponential growth of the computational complexity with the size of the strongly correlated system. At this stage, the AdS/CFT correspondence, which is also dubbed as holographic duality~\cite{Maldacena:1997re,Gubser:1998bc,Witten:1998qj,Aharony:1999ti,Zaanen:2015oix,Ammon:2015wua,Hartnoll:2018xxg,Baggioli:2019rrs}, provides a powerful framework to address this problem. It asserts that a strongly coupled quantum system in $D$ dimensions is equivalent to a weakly coupled gravity theory in $D+1$ dimensional AdS spacetime. This duality maps the complex UV dynamics of the strongly correlated boundary theory to controlled deformations of the asymptotic AdS geometry in the bulk. Therefore, the UV behavior of the system in principle may be revealed by the AdS/CFT correspondence.

In this article, we show for the first time that UV observables can probe quantum criticality with greater precision and robustness to thermal fluctuations than their IR counterparts. Specifically, we consider a holographic model characterized by metal-insulator phase transitions~\cite{Donos:2012js,Donos:2013eha,Donos:2014uba,Donos:2014oha,Ling:2014saa,Ling:2016dck,Ling:2015dma,Baggioli:2014roa,Baggioli:2016oqk,An:2020tkn,Seo:2023bdy,Ling:2016wyr}. By performing fourth-order UV asymptotic expansions and high-precision numerical simulations, we examine two types of UV observables, including high-frequency conductivity and short-range entanglement measures~\cite{Ryu:2006bv,Hubeny:2007xt,Takayanagi:2012kg,Lewkowycz:2013nqa,Nielsen:2012yss,Hayden:2011ag,Takayanagi:2017knl}. We find that the derivatives of high-frequency AC conductivity and short-range entanglement measures exhibit extrema near the quantum critical point (QCP), with critical signatures appearing as subtle variations resolvable at the sixth significant digit. These critical behaviors are analytically shown to arise from the leading deformation of the asymptotic bulk geometry. Moreover, at finite temperatures, UV observables diagnose QPTs more accurately than their IR counterparts, which enables them to serve as reliable probes of quantum criticality. Our findings establish a new paradigm for diagnosing QPTs, highlighting the unexpected richness of the UV region and its potential to complement traditional IR-focused approaches, especially at finite temperature.

	{\it Holographic MIT model.---}
To investigate this, we employ a holographic Einstein-Maxwell-Dilaton-Axion (EMDA) model, a well-established framework for studying metal-insulator transitions (MITs) driven by translational symmetry breaking~\cite{Donos:2014uba, Fu:2022qtz} (see Appendix~\ref{app:boundary_conditions} for details of the action and metric ansatz). The model is characterized by a lattice deformation with wave number $k$. By analyzing the low-energy DC conductivity, we identify a quantum critical point (QCP) at $k_c \approx 0.43$ separating a metallic phase ($k > k_c$) from an insulating phase ($k < k_c$) at zero temperature. We now turn to the UV behavior of this system near the QCP.

	{\it UV Signatures in High-Frequency Conductivity.---}
The ultraviolet (UV) behavior of the dual field theory is accessed via the high-frequency limit of the AC conductivity, $\sigma(\omega)$. In this limit, the real part asymptotically approaches $\mathrm{Re}[\sigma] \to 1$, with corrections that reveal the quantum critical signature:
\begin{equation}
	\mathrm{Re}[\sigma(\omega)] = 1 + \frac{C_\sigma}{\omega^2} + \frac{P_\sigma}{\omega^4} + \mathcal{O}(\omega^{-6}).
	\label{eq:high_frequency_UV_expansion}
\end{equation}
While the $\mathcal{O}(\omega^{-2})$ coefficient $C_\sigma$ is a $k$-independent constant, the physics of interest is encoded in the subleading coefficient $P_\sigma$, which carries the primary dependence on $k$. Isolating the variation of $P_\sigma$ is numerically demanding, as it requires extracting a small $\mathcal{O}(\omega^{-4})$ signal. This was achieved using a high-density radial grid (1000+ points) to accurately capture the oscillatory solutions, with full numerical validation presented in Appendix~\ref{app:ac_conductivity}.

\begin{figure}
	\includegraphics[width=0.475\textwidth]{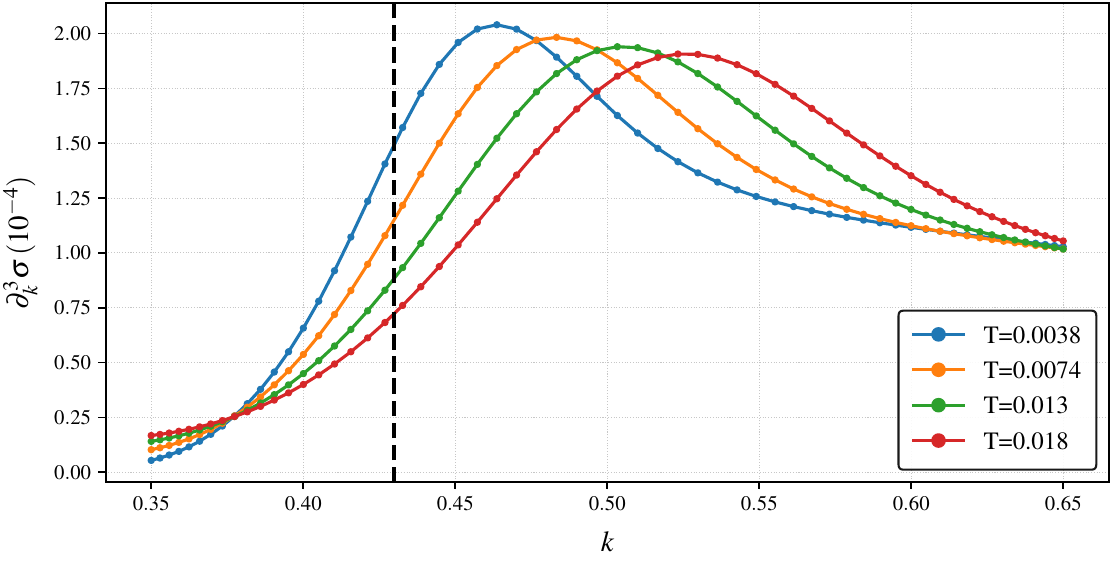}
	\caption{\textbf{UV signature of quantum criticality in high-frequency conductivity.} The third derivative of the subleading conductivity coefficient, $\partial^3_k P_\sigma$ (from Eq.~\ref{eq:high_frequency_UV_expansion}), is plotted against the lattice wave number $k$ for a fixed high frequency $\omega/\mu = 20$. As the temperature $T$ decreases, the extremum (a maximum) becomes sharper and its position approaches the zero-temperature QCP (black dashed line at $k_c \approx 0.43$), revealing a robust critical signal in the UV.}
	\label{fig:figure_simagAC_VarT}
\end{figure}

Remarkably, our numerical analysis reveals that the third derivative of the UV conductivity with respect to the parameter $k$, $\partial_k^3 \sigma_{\scriptscriptstyle{UV}}$, exhibits a pronounced extremum near the QCP, as shown in Fig.~\ref{fig:figure_simagAC_VarT}. As the temperature goes down, the location of the extremum  goes closer to the QCP. This phenomenon maintains as the frequency becomes higher. This demonstrates that the critical signature of the MIT is encoded in the high-frequency conductivity.

This observation challenges the conventional wisdom that UV physics is insensitive to quantum criticality. Conventionally, critical phenomena are understood as an IR property governed by renormalization group (RG) fixed points, with UV details being non-universal and decaying exponentially, seemingly irrelevant to the diverging correlation lengths at a QPT~\cite{Fisher:1990ux,Sachdev:2011fcc}.

This view, however, overlooks global constraints like the f-sum rule. Near a QPT, a dramatic reorganization of low-frequency (coherent) spectral weight must be compensated at high frequencies to preserve the total spectral weight, which is fixed by the system's kinetic energy. This hints that the UV region must carry signatures of the transition. Holography provides a powerful geometric explanation for this. The bulk geometry provides a geometric realization of the RG flow from the UV (AdS boundary) to the IR (horizon). A QPT forces a global reorganization of this geometry to connect different IR fixed points. This global restructuring is not confined to the deep interior; through the bulk equations of motion, it necessarily imprints itself on the near-boundary geometry, which dictates the UV observables. Crucially, the fact that the critical signature appears in the third derivative of UV observables, rather than in the observables themselves, is highly significant. It reveals that the UV does not simply see the state of the system, but rather its susceptibility—the rate at which its properties change as the critical point is approached.

This finding prompts a crucial question: Is this UV sensitivity a specific feature of transport, or a more general characteristic of quantum criticality? To explore this universality, we next examine whether similar signatures appear in geometric probes of entanglement.

\begin{figure}
	\includegraphics[width=0.5\textwidth]{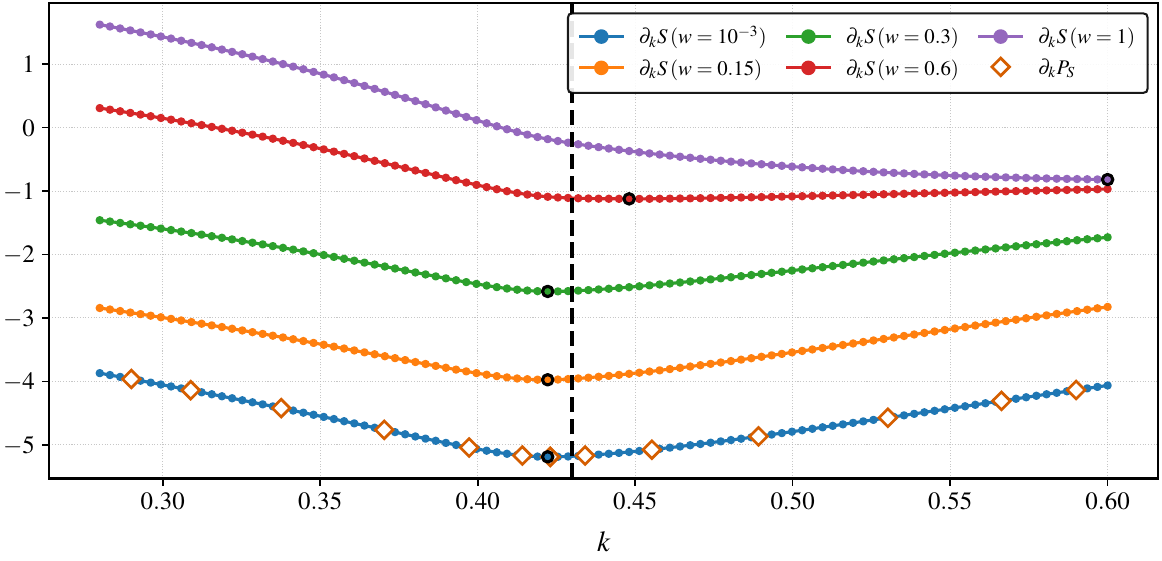}
	\caption{\textbf{Entanglement entropy as a UV probe of criticality.} The first derivative of holographic entanglement entropy (HEE), $\partial_k S$, for a strip-like region is shown as a function of $k$ for various strip widths $w$ (a UV regulator) at $T =10^{-7}$. The numerical results (solid curves) exhibit a minimum that converges to the QCP (dashed line) in the UV limit ($w \to 0$). The analytical prediction for the leading $k$-dependent term, $P_S$ (orange diamonds, see Eq.~\ref{eq:HEE_UV_expansion}), accurately reproduces the numerical behavior for small $w$.}
	\label{fig:figureofC0andSandInset}
\end{figure}

\begin{figure}
	\includegraphics[width = 0.5\textwidth]{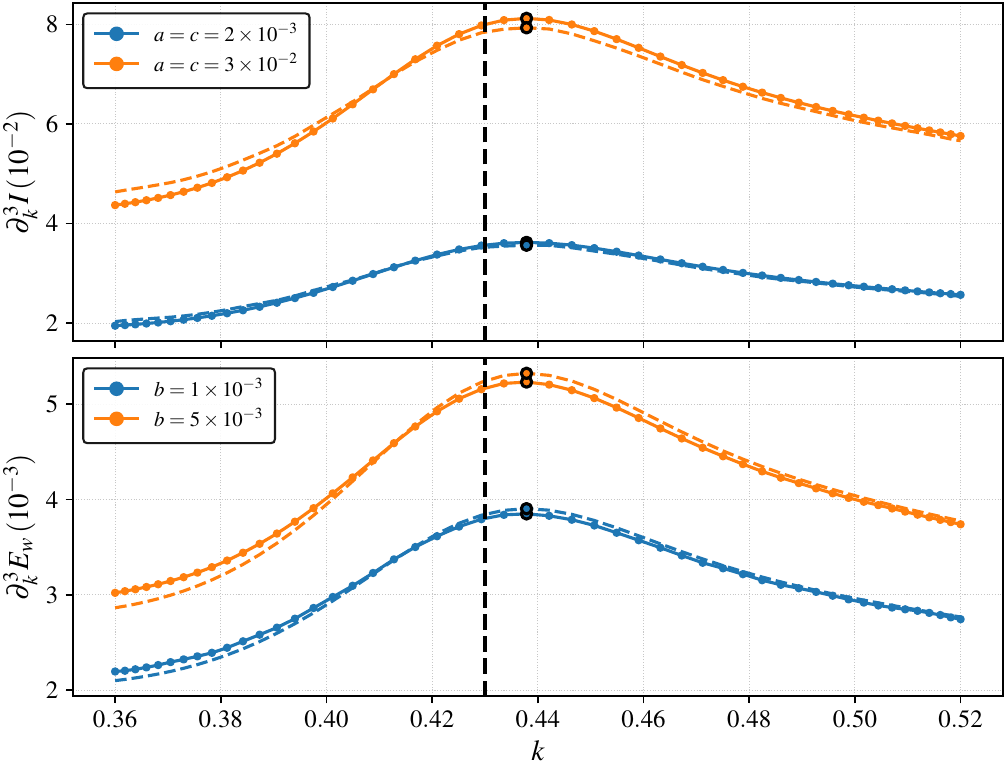}
	\caption{\textbf{UV signatures in mutual information and entanglement wedge cross-section.} The third derivatives of MI (upper panel) and EWCS (lower panel) with respect to $k$ at $T=10^{-3}$. Here, MI and EWCS are calculated for subsystems with widths $a$, $b$, and $c$. Dashed curves represent analytical results; solid curves show numerical results. Upper panel: The separation $b = 10^{-4}$ is fixed while $a$ and $c$ vary. Lower panel: $a = c = 10^{-2}$ are fixed while $b$ varies. The black dots indicate the extrema of each curve, and the black dashed line represents the critical point. The small discrepancy between analytical and numerical results arises from finite values of $a, b$, and $c$; this discrepancy diminishes as these parameters decrease, while the extremal behavior remains robust.}
	\label{fig:figureofMIandEW}
\end{figure}

{\it UV Quantum Entanglement Signatures of Criticality.---}
It is well known that quantum information measures, such as the entanglement entropy (EE), mutual information (MI) as well as reflected entropy, are traditionally used to probe long-range correlations in the IR region and thus be able to diagnose QPTs. In holography their function of diagnosing QPT has also been justified in the IR since for large subsystem configurations they exhibit extrema near the quantum critical point (QCP) \cite{Ling:2015dma,Ling:2016wyr}, as also shown in Appendix~\ref{app:holographic_info}. Notably, the seminal work by Osterloh et al.~\cite{Osterloh:2002aeb} demonstrated that the first derivative of nearest-neighbour concurrence exhibits peaks near quantum critical points, providing early evidence that short-range entanglement measures—which naturally probe UV information—can serve as sensitive diagnostics for quantum criticality. This foundational result established the active research area of entanglement-based diagnostics in condensed matter systems and provides strong theoretical support for our findings. In this section we demonstrate that these quantities surprisingly reveal sharp signatures of the QPT in the UV regime.

We investigate three key holographic entanglement measures: holographic entanglement entropy (HEE) for a strip of width $w$, MI for two disjoint regions $A$ and $C$ separated by $B$ (with widths $a, b, c$), and entanglement wedge cross-section (EWCS), a holographic measure of the reflected entropy for mixed states.

Strikingly, the first derivative of HEE with respect to the parameter $k$, $\partial_k S$, develops a distinct extremum near the QCP for strips with small widths ($w \sim 10^{-3}$ to $0.3$) (Fig.~\ref{fig:figureofC0andSandInset}). In particular, this extremal behavior becomes more evident as the width becomes smaller to approach deeper UV region. Similarly, the third derivatives of MI and EWCS, $\partial_k^3 I$ and $\partial_k^3 E_W$, also display pronounced extrema near the QCP in their UV limits (Fig.~\ref{fig:figureofMIandEW}). These observations indicate that quantum criticality is robustly imprinted in UV entanglement structures.

Analytically, we can trace the UV behavior of these measures to specific $k$-dependent coefficients in their expansions. For HEE, we find that it admits the UV expansion as $w \to 0$:
\begin{equation}
	\begin{aligned}
		S(w) =
		\frac{L_y}{4 \mu G_N} \bigg[ -\frac{C_{-1}}{w} + P_S + O(w) \bigg] ,
	\end{aligned}
    \label{eq:HEE_UV_expansion}
\end{equation}
where $C_{-1}$ remains constant and independent of $k$, while $P_S \equiv \hat V_1^{(1)}$ captures the leading $k$-dependence, where $\hat V_1^{(1)} = \partial_z V_1|_{z=0}$ represents the leading deviation of the metric component $g_{xx}$ from pure AdS$_4$ (see Appendix~\ref{app:uv_expansion} for details). The first derivative $\partial_k S$ extracts this leading $k$-dependent term and its dominant contribution is justified in  Fig.~\ref{fig:figureofC0andSandInset}, where $P_S(k)$ (orange diamond markers) accurately reproduces the numerical $\partial_k S$ for small $w$. For MI and EWCS, their dominant $k$-dependence in the UV arises from a higher-order term, $P_{IE} \equiv \hat U^{(3)} - \hat V_1^{(3)} - 2\hat V_2^{(3)}$, which captures more subtle deviations in the UV geometry (Appendix~\ref{app:uv_expansion}). We need the third derivatives $\partial_k^3 I$ and $\partial_k^3 E_W$ to isolate this term after lower-order $k$-independent pieces cancel.
The physical significance is clear: $P_S$ and $P_{IE}$ quantify how the UV geometry deforms in response to the tuning parameter $k$, encoding information about the RG flow towards different IR fixed points.

Finally, we confirm that these UV signatures are universal, holding across different model parameters $(\gamma,\,\lambda)$ within the EMDA class and in the Q-lattice model~\cite{Donos:2013eha}. This suggests they are a general feature of holographic models that exhibit quantum critical phenomena.

	{\it Robustness of UV Signatures at Finite Temperatures.---}
The above  UV sensitivity of the entanglement measures offers an exceptional advantage for diagnosing the quantum phase transition at finite temperature, since they primarily capture  quantum correlations at  short-range scales where quantum fluctuations dominate over thermal effects, while the conventional IR approach often suffers from the mixture of quantum and classical (thermal) correlations at large scales. This is evident from the AdS/CFT mapping: UV observables probe the near-boundary geometry ($z \to 0$), far from the thermal horizon at $z=1$, thus isolating quantum entanglement from thermal fluctuations that dominate at larger (IR) scales.

\begin{figure}
	\includegraphics[width = 0.475\textwidth]{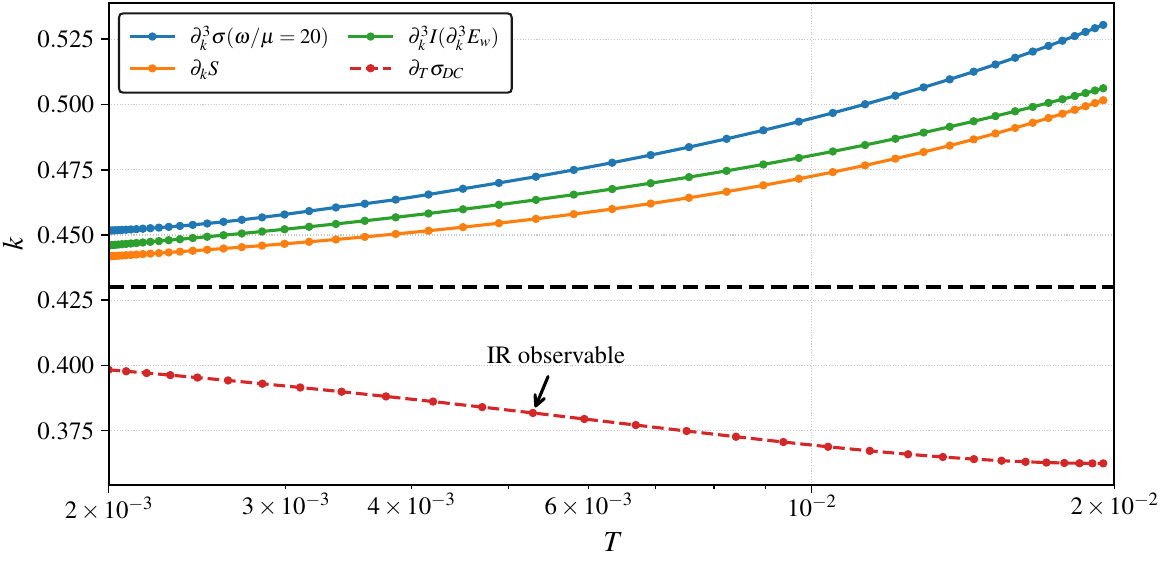}
	\caption{\textbf{Robustness of UV probes against thermal smearing.} Comparison of the estimated critical point $k_c(T)$ as a function of temperature $T$. The estimates from various UV observables (solid lines, derived from the extrema in $\partial^3_k P_\sigma$, $\partial_k S$, etc.) remain stable and close to the true zero-temperature QCP (black dashed line). In contrast, the estimate from a conventional IR probe (red dashed line, derived from the zeros of $\partial_T\sigma_{\text{DC}}$) deviates significantly as temperature increases, demonstrating the superior robustness of the UV diagnostics.}
	\label{fig:UV_temp_dependence}
\end{figure}

To justify the above observation we compare the deviation of the critical values evaluated by the DC conductivity in IR region with those evaluated by UV observables at finite temperature, as shown in Fig.~\ref{fig:UV_temp_dependence}. We find that the sharp extrema observed in the derivatives of HEE, MI, EWCS, and conductivity in the UV region persist with essentially unchanged magnitude and location across a range of temperatures. More importantly, at low temperatures (e.g., $T = 0.002$ to $0.015$) the critical values evaluated by these UV observations are closer to the zero-temperature critical point $k_{c} =0.43$ than those obtained from the DC conductivity, highlighting their robustness to thermal fluctuations and reliability in identifying quantum criticality.

{\it Summary and discussion.---}
In this article, we have demonstrated that UV observables in holographic systems can serve as potent diagnostics for QPTs, challenging the conventional wisdom that such phenomena are exclusively governed by IR physics. Table~\ref{tab:UVIR_comparison} presents a summary of the distinct physical roles and characteristics of UV and IR observables. Focusing on a holographic model with MIT, we have revealed that the UV-dominated quantities exhibit extremal behavior  near the QCP, including the third derivatives of high-frequency AC conductivity, MI, EWCS as well as the first derivative of HEE for a strip with tiny width.

These UV signatures result from the leading $k$-dependent deviations from the asymptotic AdS geometry, as shown by our analytical expansions in Appendix~\ref{app:uv_expansion}. It indicates that the UV region encodes crucial information about the underlying RG flow and the impending phase transition. A key advantage of these UV probes is their robustness at finite temperature, where they continue to provide clear diagnostics even when IR signatures might be obscured by thermal effects. This also allows for a cleaner distinction between quantum and classical correlations associated with the QPT.

The robustness of these UV diagnostics is notable. Our analytical UV expansions are largely model-independent, not relying on specific coupling parameters like $\gamma$. Numerically, the extremal UV behavior near the QCP persists for different choices of $\gamma$ and $\lambda$. Crucially, these UV signatures of criticality are not confined to the EMDA model; we have verified their presence in other frameworks, such as the Q-lattice model \cite{Donos:2013eha}, indicating they represent a general feature of the UV region in holographic QPTs.

Our findings establish a new paradigm for investigating QPTs, shifting focus towards the rich, yet often overlooked, information content of the UV sector. This paradigm fundamentally reshapes our understanding on the interplay between UV and IR dynamics in holographic systems and quantum criticality. The ability to detect QPTs from short-distance physics opens new avenues for theoretical exploration, potentially extending to topological QPTs or systems beyond the Landau-Ginzburg framework. Furthermore, it may inspire novel experimental probes, such as high-harmonic generation spectroscopy or precision measurements of short-range correlations, to search for similar UV signatures of quantum criticality in strongly correlated materials like cuprates or heavy-fermion systems. We anticipate that this IR-UV dual diagnostic framework could be extended to a wider class of physical quantities and models, offering a more complete picture of critical phenomena.

\begin{table}[htbp]
	\renewcommand{\arraystretch}{1.5}
	\caption{Comparison between IR and UV observables.}
	\begin{tabular}{m{4.cm}<{\centering}|m{4.cm}<{\centering}}
		\hline
		IR observable                         & \textbf{UV observable}                                 \\
		\hline
		Probe low-energy excitations          & \textbf{Probe high-energy degrees of freedom}          \\
		Capture long-range correlations       & \textbf{Capture short-range correlations}              \\
		Limit: Smearing of the thermal signal & \textbf{Advantage: Distinction of the critical signal} \\
		\hline
	\end{tabular}
	\label{tab:UVIR_comparison}
\end{table}

{\it Data Availability.---} The core numerical data and analysis scripts required to reproduce the key findings of this paper are publicly available at our GitHub repository \url{https://github.com/physicsuniverse/EMDA-UV-QPT}. We welcome the community to test and build upon our results.

	{\it Acknowledgement.---} Peng Liu would like to thank Yun-Ha Zha and Yi-Er Liu for their kind encouragement during this work. This work is supported by the Natural Science Foundation of China under Grant Nos. 12475054, 12035016, 12275275, 12375055 and the Guangdong Basic and Applied Basic Research Foundation No. 2025A1515012063. Zhe Yang is supported by the Jiangsu Postgraduate Research and Practice Innovation Program under Grant No. KYCX25\_3922.

\clearpage
\onecolumngrid

\appendix
\section{EMDA Model Details and Phase diagram}
\label{app:boundary_conditions}

The four-dimensional Einstein-Maxwell-Dilaton-Axion (EMDA) model is characterized by the Lagrangian:
\begin{equation}
	\begin{split}
		\mathcal{L} = & R + 6\, \cosh\,\psi - \frac{3}{2}\left[ \left(\partial \psi \right)^2 +4 \, \sinh^2 \psi \left( \partial \chi \right)^2 \right]  - \frac{1}{4} \, \cosh^{\gamma/3} \left( 3\psi \right) F^2 ,
	\end{split}
	\label{lang}
\end{equation}
where $R$ is the Ricci scalar, and $F=dA$ is the Maxwell field strength (with $A$ being the Maxwell field). $\psi$ is a scalar dilaton field, and $\chi$ is an axion field. The presence of the axion, specifically chosen as $\chi= \bar{k} x$, explicitly breaks translation symmetry in the $x$-direction, mimicking the effect of a lattice in condensed matter system. The parameter $\gamma$ governs the coupling between the dilaton and the Maxwell field, playing a crucial role in determining the nature of the IR fixed points.

We seek planar black brane solutions with the following ansatz:
\begin{equation}
	\begin{aligned}
		 & ds^2=\frac{1}{z^2}\left( -P dt^2+\frac{dz^2}{P}+V_1dx^2+V_2dy^2 \right), \\
		 & P = U(1-z) \left( 1+z+z^2-\frac{\mu^2 z^3}{4} \right),                   \\
		 & A=\mu(1-z)adt,\qquad \chi=\bar{k} x,\qquad \psi = z^{3-\Delta} \phi,
		\label{metric}
	\end{aligned}
\end{equation}
where $\mu$ is the chemical potential of the dual field theory, and $\Delta=2$ is the conformal dimension of the scalar operator dual to $\psi$. In this coordinate system, the AdS boundary is located at $z=0$, and the black brane horizon is at $z=1$. The functions $U$, $V_1$, $V_2$, $a$, and $\phi$ are functions of the radial coordinate $z$ only.

Three dimensionless parameters characterize the solutions: the normalized Hawking temperature $T \equiv \bar{T}/\mu$, the normalized lattice amplitude $\lambda \equiv \bar{\lambda}/\mu$, and the normalized wave number $k \equiv \bar{k}/\mu$. Here, $\bar{\lambda}$ corresponds to the boundary value of the dilaton field $\psi(0)$, effectively setting the strength of the lattice deformation. The chemical potential $\mu$ serves as the scaling unit, placing the system in a grand canonical ensemble, and the Hawking temperature follows the relation
\begin{equation}
	\bar T = \frac{12-\mu^2}{16 \pi} U(1).
\end{equation}
To obtain the background solution numerically, we impose the following UV boundary conditions:
\begin{equation}
	U(0)=1, \, V_1(0)=1,\, V_2(0)=1,\, a(0)=1,\, \phi(0) = \bar{\lambda}.
	\label{UV_boundary_conditions}
\end{equation}

The EMDA model exhibits rich phase structure at zero temperature, where varying $\lambda$ or $k$ can drive distinct QPTs by altering the nature of the IR fixed points. Specifically, the IR fixed points can be classified by the value of $\gamma$ under the linear perturbations of the near horizon geometry, which is $\text{AdS}_2 \times \mathbb{R}^2$~\cite{Donos:2014uba,Fu:2022qtz,Gong:2023tbg}. Correspondingly, the conductivity of the dual system at different IR fixed points may exhibit metallic or insulating behavior.

\begin{figure}
	\centering
	\includegraphics[width=0.5\textwidth]{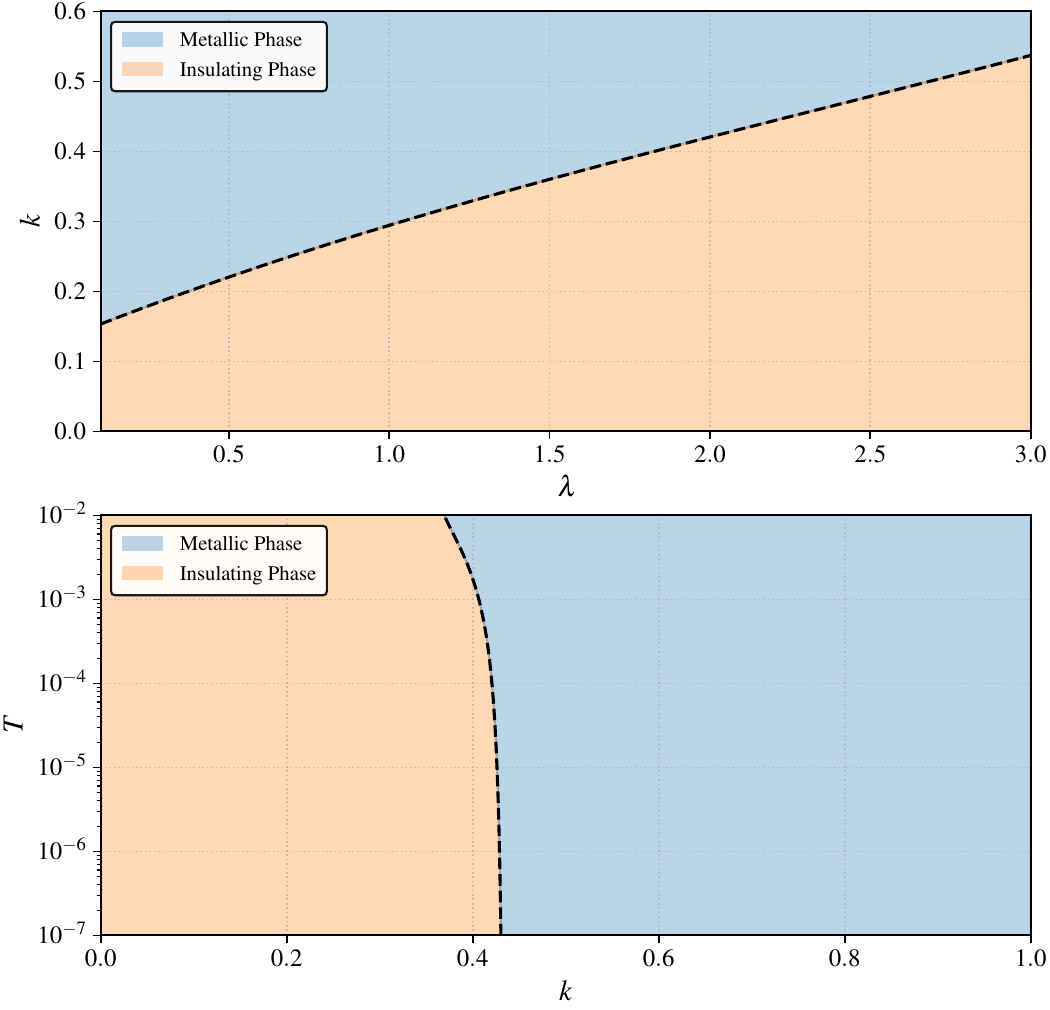}
	\caption{The phase diagram of the EMDA model, the black dashed line represents the critical point. Upper panel: the phase diagram of $\lambda-k$ at $T=10^{-4}$. Lower panel: the phase diagram of $k-T$ at $\lambda=2$. The critical point shows stable behavior with the decreasing of the temperature.}
	\label{fig:phasediagram}
\end{figure}

To distinguish metallic and insulating phases, we examine the temperature derivative of DC conductivity in the zero temperature limit: $\partial_T \sigma_{DC}\big|_{T \to 0} < 0$ (metallic), $\partial_T \sigma_{DC}\big|_{T \to 0} > 0$ (insulating), and $\partial_T \sigma_{DC}\big|_{T \to 0} = 0$ (QCP). This allows us to construct the phase diagram in the $\lambda$-$k$ plane (see in Fig. \ref{fig:phasediagram}).

Without loss of generality, we fix the dilaton-Maxwell coupling exponent to $\gamma = -1/6$ and the lattice amplitude to $\lambda = 2$. In Fig. \ref{fig:phasediagram}, we approach the zero-temperature limit by setting the numerical temperature to $T = 10^{-7}$. Under these conditions, we find that the QCP is located at $k_{c} = 0.43$, with the metallic phase for $k > k_{c}$ and the insulating phase for $k < k_{c}$.

\section{Details of the AC Conductivity Calculation}
\label{app:ac_conductivity}
To compute the AC conductivity along the direction of lattice, i.e., $x$-direction, we introduce linear perturbations of the form
\begin{equation}
	g_{tx} = e^{-i \omega t} \delta h_{tx}(z), \quad A_{x} = e^{-i \omega t} \delta a_{x}(z), \quad \chi = e^{-i \omega t} \delta \chi(z),
\end{equation}

\begin{figure}
	\includegraphics[width=0.5\textwidth]{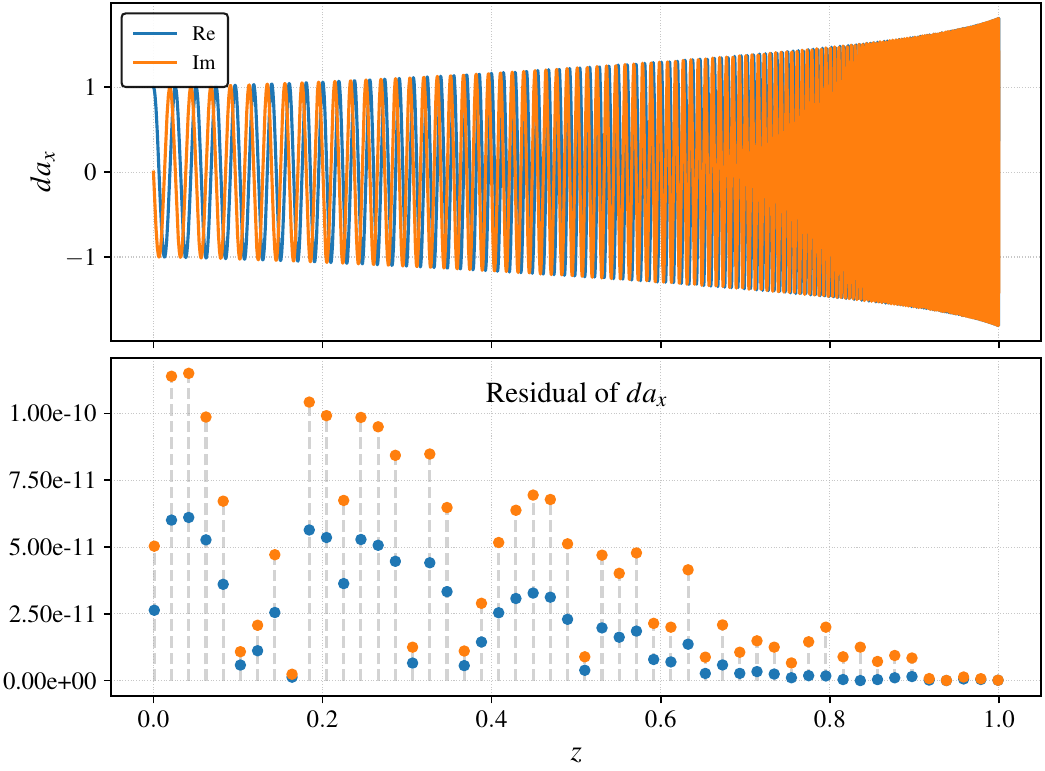}
	\caption{The schematic of the solution $d a_x$ of the high-frequency AC conductivity at $\omega/\mu=20$ and $T=0.005$. Upper panel: The solution of $d a_x$ and it shows rapid oscillations. Lower panel: The residual of the solution $d a_x$, which demonstrate our solution is highly reliable.}
	\label{fig:figure_ax}
\end{figure}
leading to a set of coupled linear differential equations for $\{ \delta h_{tx}(z), \delta a_x(z), \delta \chi(z)\}$. At the horizon $z=1$, we impose regular and ingoing boundary conditions
\begin{equation}
	\begin{aligned}
		\delta h_{tx}(z) & = (1-z)^{1- \frac{i \omega}{4 \pi \mu T}} dh_{tx}(z), \\
		\delta a_{x}(z)  & =  (1-z)^{- \frac{i \omega}{4 \pi \mu T}} da_{x}(z),  \\
		\delta \chi(z)   & =  (1-z)^{- \frac{i \omega}{4 \pi \mu T}} d\chi(z),
	\end{aligned}
\end{equation}
where $ dh_{tx}(z)$, $da_x(z)$, and $d\chi(z) $ are regular at the horizon.

At the UV boundary $z=0$, we set $da_x(0)=1$ to define the source of the external electric field. Additionally, to ensure that the response corresponds to a current-current correlator in the dual field theory, we impose the constraint
\begin{equation}
	d\chi(0) - \frac{i k}{\omega} dh_{tx}(0) = 0,
\end{equation}
which comes from the diffeomorphism and gauge invariance.

Solving these equations numerically yields the perturbations $\{ \delta a_x(z),\delta h_{tx}(z), \delta \chi(z) \}$, and finally, the AC conductivity is extracted via the Kubo formula:
\begin{equation}
	\sigma(\omega) = \frac{ \delta a_x'(z) }{ i \omega \delta a_x(z) } \Bigg|_{z \to 0}.
\end{equation}
Our results show robust numerical convergence. All numerical calculations have been carefully validated and double-checked to ensure the reliability of our findings.

Two main numerical challenges arise when we evaluate $\sigma_\text{AC}$ at low temperatures and high frequencies:
\begin{enumerate}
	\item Rapid oscillations: The perturbations $\{ da_x(z), dh_{tx}(z), d\chi(z) \}$ exhibit strong oscillatory behavior near the horizon. This requires a high spatial resolution along the radial direction $z$ to accurately capture the rapid oscillatory behavior of perturbations, as shown in Fig.~\ref{fig:figure_ax}. For example, in the case $T=0.005$ and $\omega/ \mu = 20$, we discretize the domain using $1000$ grid points in the $z-$direction to ensure numerical stability and convergence. Since the oscillation scale is controlled by the ratio $ \omega / T $, increasing $ \omega / T $ demands higher spatial resolution.
	\item Small signal strength: In the UV limit, the real part of the AC conductivity approaches $1$ asymptotically. The critical signal arises from subtle $k$-dependent deviations from this limiting value. As shown in Table~\ref{tab:UVReSigma}, for fixed values of $(T,\,\omega/\mu)$, the variation in $\mathrm{Re}[\sigma_\text{AC}]$ across the interval $k \in [0.38, 0.53]$ is extremely small—on the order of $10^{-6}$ to $10^{-8}$—requiring high-precision numerics to identify critical signal. This subtle signal stems from its appearance at the order $\omega^{-4}$ in the high-frequency expansion, as shown in Eq.~\ref{eq:high_frequency_UV_expansion}.
\end{enumerate}

\begin{table}[h]
	\centering
	\caption{Real part of the high-frequency conductivity $\mathrm{Re}[\sigma_\text{AC}]$ for varying $k$ at different $(T,\omega/\mu)$. For clarity, only five representative $k$-values are shown here. In our calculations, the full interval $k \in [0.33, 0.53]$ is discretized with 54 points.}
	\renewcommand{\arraystretch}{1.2}
	\begin{tabular}{c|ccccc}
		\hline\hline
		                   & \multicolumn{5}{c}{$k$}                                                     \\
		\cline{2-6}
		$(T,\ \omega/\mu)$ & 0.33                    & 0.38       & 0.43       & 0.48       & 0.53       \\
		\hline\hline
		(0.01,\ 20)        & 1.00126150              & 1.00125966 & 1.00125760 & 1.00125533 & 1.00125267 \\
		(0.002,\ 20)       & 1.00126148              & 1.00125963 & 1.00125756 & 1.00125528 & 1.00125262 \\
		(0.01,\ 50)        & 1.00020052              & 1.00020050 & 1.00020047 & 1.00020045 & 1.00020041 \\
		\hline\hline
	\end{tabular}
	\label{tab:UVReSigma}
\end{table}

\section{Holography quantum information}
\label{app:holographic_info}
Entanglement entropy (EE) is a key measure of quantum entanglement, particularly suitable for pure states. The area of the minimal surface in the bulk gives its holographic counterpart, HEE \cite{Ryu:2006bv, Hubeny:2007xt, Takayanagi:2012kg, Lewkowycz:2013nqa}:
\begin{equation}
	S_A = \frac{\text{Area}(\gamma_A)}{4G_N},
\end{equation}
where $G_N$ is Newton's constant, and $\gamma_A$ is the minimal surface.

\begin{figure}
	\includegraphics[height=0.13\textheight]{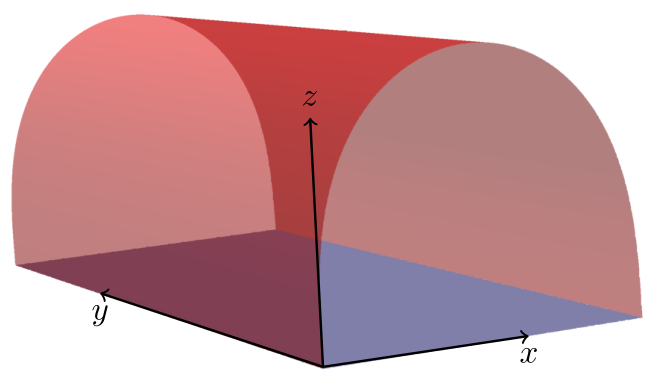}
	\includegraphics[height=0.13\textheight]{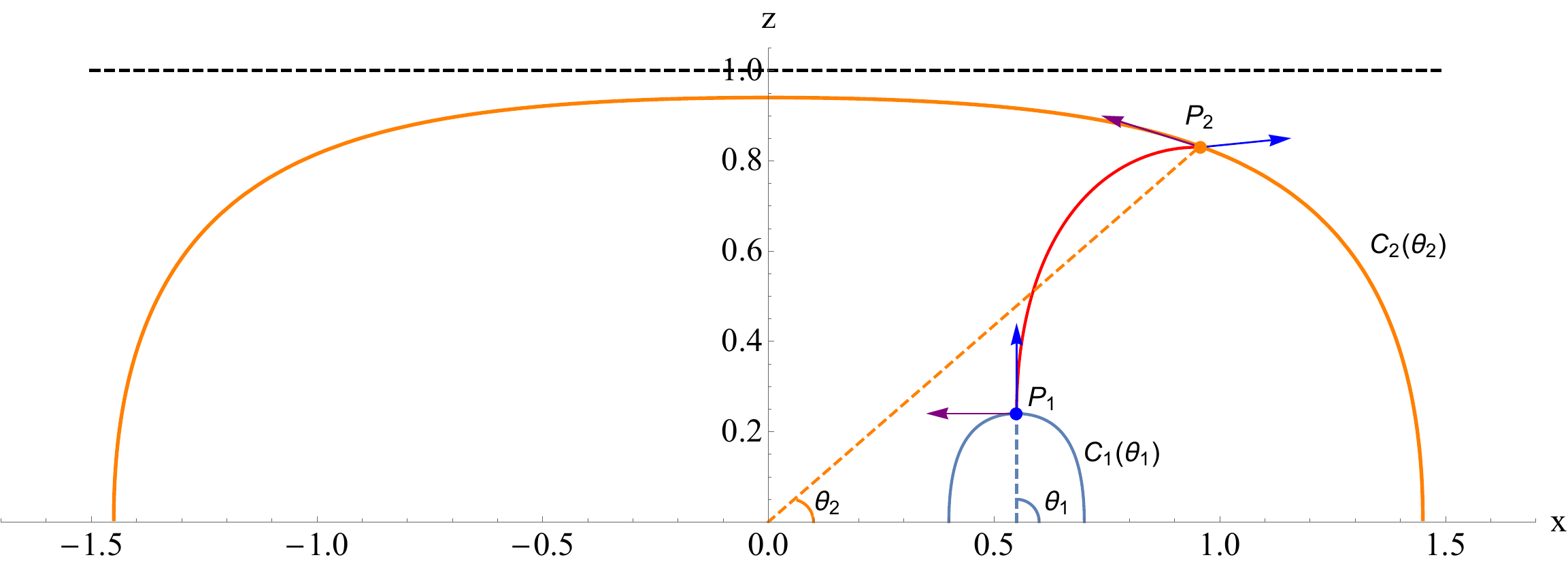}
	\caption{Left: Schematic plot of HEE, where the red surface represents the geometry of HEE. Right: Schematic plot of EWCS in $x-z$ plane. The red curve shows the minimum cross-section between minimum surfaces $C_1$ and $C_2$ of the entanglement wedge. The blue and purple arrows represent tangent vectors $\left( \frac{\partial}{\partial z} \right)^a$ and $\left( \frac{\partial}{\partial \theta} \right)^a$, respectively. The local perpendicular condition of these tangent vectors determines the position of EWCS. The dark horizontal line represents the black brane horizon.}
	\label{fig:sy}
\end{figure}

We examine a strip that is infinitely long in the $y$-direction and has a finite width $w$ in the $x$-direction, as depicted in Fig. \ref{fig:sy}. This allows us to focus on the $x$-direction where $\chi$ breaks translational symmetry. The HEE is characterized by the maximum depth $z_s$ of the minimal surface. For the metric in Eq.~\eqref{metric}, $w$ and the HEE $S$ are related to $z_s$ as:
\begin{equation}
	\begin{aligned}
		w & = 2 \mu \int_0^{z_s} dz \, z^2 \sqrt{\frac{V_1(z_s) V_2(z_s)}{P(z) V_1(z) W(z_s, z)}},                                                                           \\
		S & = \frac{L_y}{2 \mu G_N} \left[ -\frac{1}{z_s} + \int_0^{z_s} \frac{dz}{z^2} \left( \frac{z_s^2 V_1(z) V_2(z)}{\sqrt{P(z) V_1(z) W(z_s, z)}} - 1 \right) \right],
		\label{Calculation equation for S and w under arbitrary zs}
	\end{aligned}
\end{equation}
where $W(z_s,z) \equiv z_s^4 V_1(z) V_2(z) - z^4 V_1(z_s) V_2(z_s)$, and $L_y(\to \infty)$ is the length of the infinitely long strip in the $y$ direction.

Quantum states in real systems are typically mixed states, for which HEE is insufficient as an entanglement measure. To quantify entanglement in mixed states, MI has been proposed. For disjoint regions $A$ and $C$ separated by $B$, MI is defined as \cite{Nielsen:2012yss, Hayden:2011ag}:
\begin{equation}
	I(A;\,C) = S(A) + S(C) - S(A \cup C),
	\label{eq:definition_of_I}
\end{equation}
where $S(X)$ is the HEE of region $X$. The HEE of the combined region $A \cup C$ is given by:
\begin{equation}
	S(A \cup C) = \min\{S(A)+S(C), \,S(B)+S(A+B+C)\}.
	\label{eq:definition_of_S_AC}
\end{equation}
MI provides a clear indicator of entanglement: $I(A; \,C) = 0$ for non-entangled regions, and $I(A; \,C) > 0$ for entangled regions.

Another proposed measure for mixed state entanglement is the EWCS. The EWCS is defined as the area of the minimum cross-section within the entanglement wedge \cite{Takayanagi:2017knl}:
$E_W(\rho_{AB}) = \min_{\Sigma_{AB}} \left( \frac{\text{Area}(\Sigma_{AB})}{4G_N} \right).$
The EWCS is calculated as:
\begin{equation}
	E_W = \frac{L_y}{4 \mu G_N} \int_{C_{p_1,p_2}}\sqrt{g_{xx}g_{yy}x'(z)^2+g_{zz}g_{yy}}dz,
	\label{eq:EWCS}
\end{equation}
with the corresponding equation of motion:
\begin{equation}
	\begin{aligned}
		x'(z)^3 \left( \frac{g_{xx}g'_{yy}}{2g_{yy}g_{zz}}+\frac{g'_{xx}}{2g_{zz}}\right) +x'(z)\left( \frac{g'_{xx}}{g_{xx}}+\frac{g'_{yy}}{2g_{yy}}-\frac{g'_{zz}}{2g_{zz}} \right) +x''(z)=0,
	\end{aligned}
	\label{eq:EOM_of_EWCS}
\end{equation}
where primes denote derivatives with respect to $z$, and $C_{p_1,p_2}$ is the global minimal curve connecting the minimum surfaces $C_1$ and $C_2$, as depicted in Fig. \ref{fig:sy}. For HEE calculations, the configuration parameter $w$ represents the width of the single region under consideration. For MI and EWCS calculations, the configuration parameters $(a, b, c)$ represent the widths of regions $A$, $B$, and $C$, respectively.

\begin{figure}
	\includegraphics[width = 0.5\textwidth]{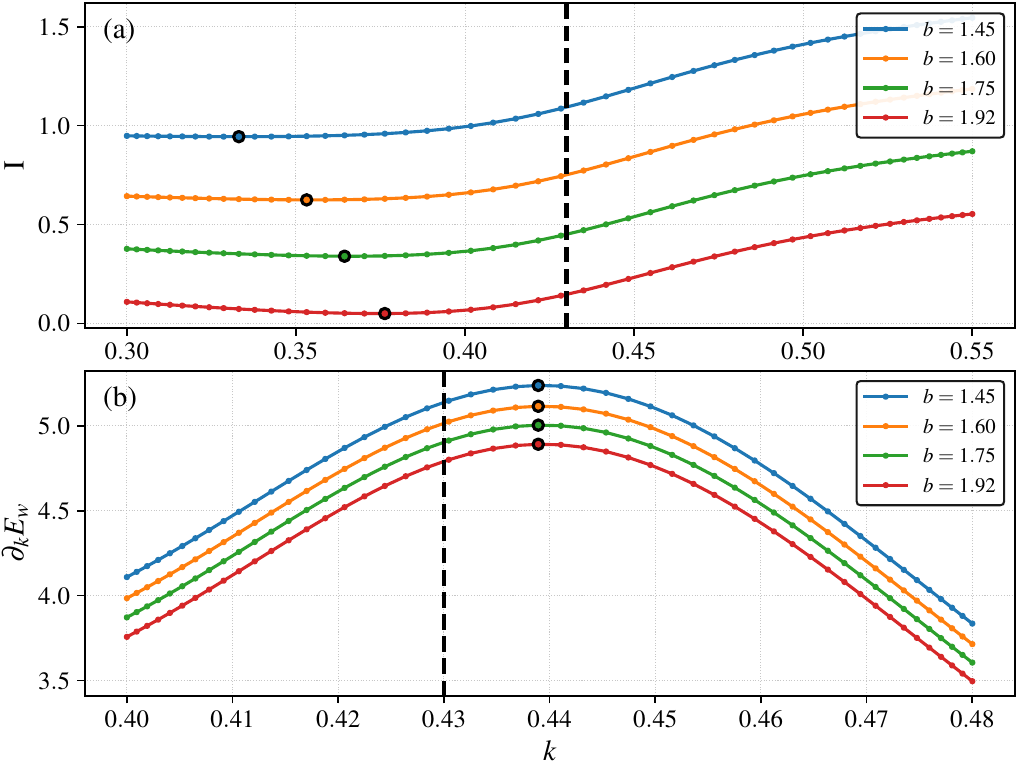}
	\caption{The schematic of MI (a) and first derivative of EWCS (b) with respect to $k$ for $a = c = 40$, with $b$ varied. The black dots indicate the extrema of each curve, and the black dashed line represents the critical point.}
	\label{fig:figure_MIandEW_IR}
\end{figure}

In addition to UV observables, we also examine the MI and EWCS in the IR region, both of which exhibit extrema near the QCP, as shown in Fig.~\ref{fig:figure_MIandEW_IR}.

\section{UV Series Expansion of Quantum Entanglement Measures}
\label{app:uv_expansion}

Here we detail the UV series expansion for HEE, MI, and EWCS. For HEE, we consider a strip of width $w$ and length $L_y$ (taken to infinity). The HEE $S$ and width $w$ are determined by the turning point $z_s$ of the minimal surface in the bulk, as given by Eq.~\eqref{Calculation equation for S and w under arbitrary zs}.

In the UV limit, $w \to 0$, which corresponds to $z_s \to 0$. Expanding $w(z_s)$ and $S(z_s)$ from Eq.~\eqref{Calculation equation for S and w under arbitrary zs} in powers of $z_s$:
\begin{equation}
	\begin{aligned}
		w & = 2 \mu \left[ \Gamma z_s + f_1 z_s^2 + f_2 z_s^3 + f_3 z_s^4 + O(z_s^5) \right],                    \\
		S & = \frac{L_y}{ 2 \mu G_N } \left[ -\frac{\Gamma}{z_s} + g_1 + g_2 z_s + g_3 z_s^2 + O(z_s^3) \right],
	\end{aligned}
	\label{eq:app_l_and_S_in_zs_to_0}
\end{equation}
where $\Gamma = \frac{\sqrt{\pi} \Gamma(3/4) }{ \Gamma(1/4)}$ is a constant. The coefficients $f_i$ and $g_i$ are functions of the near-boundary values of the metric and scalar field components and their $z$-derivatives at $z=0$: $\hat U^{(n)} \equiv \partial_z^n U|_{z=0}$, $\hat V_1^{(n)} \equiv \partial_z^n V_1|_{z=0}$, etc. These are related by the equations of motion at the boundary, e.g.,
\begin{equation}
	\begin{aligned}
		 & \hat{U}^{(1)} = \hat{V}_1^{(1)} = \hat{V}_2^{(1)},  \quad \hat{U}^{(2)} = \hat{V}_1^{(2)} = \hat{V}_2^{(2)},                     \\
		 & \left( \hat{U}^{(1)} \right)^2 - 2 \hat{U}^{(2)} = 3 \left( \hat{\phi}^{(0)} \right)^2,                                          \\
		 & \hat{V}_2^{(3)} + \hat{V}_1^{(3)} = -6 \left( \hat{\phi}^{(0)} \right)^2 \hat{V}_1^{(1)} - 12 \hat{\phi}^{(0)} \hat{\phi}^{(1)},
	\end{aligned}
	\label{eq:app_boundary_field_conditions}
\end{equation}
where $\hat \phi^{(0)} = \mu \lambda$. The explicit forms of Eq.~\eqref{eq:app_l_and_S_in_zs_to_0} are:
\begin{equation}
	\begin{aligned}
		 & \lim_{z_s \to 0} w = 2 \mu \Bigg[ \Gamma z_s
		- \frac{1}{2} \hat{V}_1^{(1)} \Gamma z_s^2
		+ \left( \frac{\Gamma}{4} \left( \hat{V}_1^{(1)} \right)^2
		- \frac{\pi - 12 \Gamma^2}{32 \Gamma} \mu^2 \lambda^2 \right) z_s^3       \\
		 & + \left( \frac{(4 + \mu^2) \pi}{64}
		+ \frac{3 \pi - 8 \pi \Gamma - 20 \Gamma^2}{64 \Gamma} \mu^2 \lambda^2 \hat{V}_1^{(1)}
		- \frac{\Gamma}{8} \left( \hat{V}_1^{(1)} \right)^3
		- \frac{\pi - 2 \Gamma}{4} \mu \lambda \hat{\phi}^{(1)}
		- \frac{\pi}{96} (\hat{U}^{(3)} + \hat{V}_1^{(3)}) \right) z_s^4
		+ O(z_s^5) \Bigg],                                                        \\
		 & \lim_{z_s \to 0} S = \frac{L_y}{ 2 \mu G_N} \Bigg[ -\frac{\Gamma}{z_s}
		+ \frac{1 - \Gamma}{2} \hat{V}_1^{(1)}
		+ \frac{3 (4 \Gamma^2 - \pi)}{32 \Gamma} \mu^2 \lambda^2 z_s              \\
		 & + \Bigg( \frac{\pi (4 + \mu^2)}{32}
		+ \frac{3 \pi - 16 \pi \Gamma + 4 \Gamma^2}{64 \Gamma} \mu^2 \lambda^2 \hat{V}_1^{(1)}
		+ \frac{ \Gamma - \pi }{2} \mu \lambda \hat{\phi}^{(1)}
		- \frac{\pi}{48} \left( \hat{U}^{(3)} + \hat{V}_1^{(3)} \right) \Bigg) z_s^2
		+ O(z_s^3) \Bigg].
	\end{aligned}
	\label{eq:relationship_S_l_zs_small_width}
\end{equation}
Inverting $w(z_s)$ to find $z_s(w)$:
\begin{equation}
	\begin{aligned}
		 & z_s = \frac{w}{2 \Gamma \mu} + \frac{ \hat V_1^{(1)} w^2 }{8 \Gamma^2 \mu^2} + \left[ (\pi-12\Gamma^2)\mu^2 \lambda^2 +  8 \Gamma^2 \left( \hat V_1^{(1)} \right)^2 \right] \frac{w^3}{256 \Gamma^5 \mu^3} + \bigg[ - 3 \Gamma \pi (4+\mu^2)
		\\
		 & + 6 \left( \pi + 4 \pi \Gamma - 20 \Gamma^2 \right) \mu^2 \lambda^2 \hat V_1^{(1)} + 24 \Gamma^2 \left( \hat V_1^{(1)} \right)^3 + 48 \Gamma (\pi - 2\Gamma)\mu \lambda \hat \phi^{(1)} + 2 \pi \Gamma \left( \hat U^{(3)} + \hat V_1^{(3)}\right) \bigg] \frac{w^4}{3072 \Gamma^6 \mu^4} + O(w^5).
		\label{eq:zs_small_w}
	\end{aligned}
\end{equation}
Substituting $z_s(w)$ into $S(z_s)$ yields $S(w)$ for small $w$:
\begin{equation}
	\begin{aligned}
		S(w) =
		\frac{L_y}{4 \mu G_N} \bigg[ -\frac{4 \Gamma^2 \mu}{w} + \underbrace{\hat V_1^{(1)}}_{P_S} - \frac{\pi \lambda^2 \mu}{16 \Gamma^2} w
		+ \left( C_{21} + C_{22} \underbrace{(\hat U^{(3)} - \hat V_1^{(3)} - 2\hat V_2^{(3)})}_{P_{IE}} \right) w^2 + O(w^3) \bigg] ,
		\label{eq:app_S_of_w_in_small_w}
	\end{aligned}
\end{equation}
where $C_{21} = \frac{\pi(4+\mu^2)}{128 \Gamma^2 \mu^2}$ and $C_{22} = - \frac{\pi} {192 \Gamma^2 \mu^2}$ are constants independent of $k$. The $k$-dependence of $S(w)$ at leading order (constant term in $w$) comes from $P_S \equiv \hat V_1^{(1)}$. The next significant $k$-dependent term appears at $O(w^2)$ through $ P_{IE} \equiv \hat U^{(3)} - \hat V_1^{(3)} - 2\hat V_2^{(3)} $.

For MI and EWCS, we consider a symmetric configuration $a=c=n\,b$ (for some constant $n$) and take $b \to 0$. In the presence of entanglement, combining Eqs.~\eqref{eq:definition_of_I}, \eqref{eq:definition_of_S_AC} and the UV expansion of $S$ \eqref{eq:app_S_of_w_in_small_w} yields:
\begin{equation}
	\begin{aligned}
		I = S(nb) + S(nb) - S((2n+1)b) - S(b) = \frac{L_y}{ 4 \mu G_N } \bigg[ C_{I0} \frac{1}{b} + C_{I1} b
			+ \left(C_{I21} + C_{I22} P_{IE} \right) b^2 + O(b^3) \bigg] ,
		\label{eq:app_MI_expression}
	\end{aligned}
\end{equation}
where $C_{I0}, C_{I1}, C_{I21}$, and $C_{I22}$ are $k$-independent coefficients derived from the $w^{-1}, w^1$, and $w^2$ terms in Eq.~\eqref{eq:app_S_of_w_in_small_w}. Importantly, the $P_S$ term cancels out in MI, and the leading $k$-dependence of MI comes from $ P_{IE}$ at $O(b^2) $.

For EWCS, in the symmetric configuration $a=c$, the minimal cross-section $\Sigma_{AB}$ is a surface at constant $x$ that extends from $z=0$ to some $z_s = z_{max}$). In the UV limit ($b \to 0$), the dominant contribution comes from $z \approx 0$. A similar expansion for $E_W$ (Eq.~\eqref{eq:EWCS}) yields:
\begin{equation}
	\begin{aligned}
		E_W = \frac{L_y}{4 \mu G_N} \bigg[ C_{E0} \frac{1}{b} + C_{E1} b + \left( C_{E21} + C_{E22} P_{IE} \right) b^2 + O(b^3) \bigg] .
	\end{aligned}
	\label{eq:app_EWCS_expression}
\end{equation}
Again, $C_{E0}, C_{E1}, C_{E21}$, and $C_{E22}$ are $k$-independent, and the leading $k$-dependence arises from $ P_{IE}$ at $O(b^2)$. The precise coefficients $C_{I0}, C_{I1}, C_{I21}, C_{I22}$ and $C_{E0}, C_{E1}, C_{E21}, C_{E22}$ depend on $n$ and other constants but their $k$-independence is what matters.

Thus, in the UV limit, both MI and EWCS have their leading $k$-dependence governed by the same geometric deformation term $ P_{IE} \equiv \hat U^{(3)} - \hat V_1^{(3)} - 2\hat V_2^{(3)} $:
\begin{equation}
	\partial_k^n I \propto \partial_k^n E_W \propto \partial_k^n P_{IE}.
	\label{eq:app_MI_proportional_EWCS}
\end{equation}
This explains why their third derivatives with respect to $k$ exhibit similar extremal behavior near the QCP, as shown in Fig.~\ref{fig:figureofMIandEW}.

\end{document}